\documentclass[aps,twocolumn,prb,showpacs]{revtex4}
\usepackage{graphicx}
\usepackage[]{epsfig}
\usepackage[dvips]{color}
\def\nab{{\mbox{\boldmath{$\nabla$}}}}
\def\kap{{\mbox{\boldmath{$\kappa$}}}}

\def\xiv{{\mbox{\boldmath{$\xi$}}}}

\begin{document}

\title{Organized Current Patterns in Disordered Conductors}
\author{Y. Japha$^{1}$, O. Entin-Wohlman$^{1}$, T. David$^{1}$, R. Salem$^{1}$, \\
S. Aigner$^{2,3}$, J. Schmiedmayer$^{2,3}$, and R. Folman$^{1}$ \\
$^1$Department of Physics, Ben-Gurion University of the Negev,
P.O. Box 653, Be'er-Sheva 84105, Israel.\\
$^{2}$Physikalisches Institut, Universit\"at Heidelberg, Philosophenweg 12, 69120 Heidelberg.\\
$^{3}$Atominstitut der Österreichischen Universitäten, TU-Wien, Stadionalle 2, 1020 Vienna, Austria.
}
\date{\today}

\begin{abstract}

We present a general theory of current deviations in straight current
carrying wires with random imperfections, which quantitatively
explains the recent observations of organized patterns of magnetic
field corrugations above micron-scale evaporated wires. These
patterns originate from the most efficient electron scattering by
Fourier components of the wire imperfections with wavefronts along 
the $\pm 45^{\circ}$ direction.  We show that long range effects of
surface or bulk corrugations are suppressed for narrow wires or
wires having an electrically anisotropic resistivity.

\end{abstract}

\maketitle

Electron scattering by microscopic structural imperfections in thin conducting films
is a major factor determining their  conductivity properties, especially at low 
temperatures \cite{conductivity1,conductivity2}. 
Orginary polycrystalline metal wires with straight boundaries 
are usually considered to have ohmic conductance with a homogeneous current flow 
on a scale much larger than their grain size (typically tens of nanometers)
\cite{random}. 
Measurements using ultracold atoms as a highly sensitive probe to 
minute changes in the magnetic field have revealed directional deviations of 
the current flow far from the edges of the wire \cite{schmiedmayer_nature}. 
Recent observations of atomic density fluctuations a few microns above wires of
different thickness and grain size have revealed organized patterns of 
current flow directional deviations which are oriented predominantly at $\pm 45^{\circ}$
relative to the wire axis, on a length scale as large as tens of microns \cite{simon}. 
It was shown that this effect is a general property 
of electron scattering by random imperfections in the conductor.
In contrast to previous observations of atomic density fluctuations above
current carrying wires \cite{tubingen, keterle, jones, hd frag, orsay1}, 
which were attributed to current irregularities due to wire edge corrugations 
\cite{orsay1, lukin, orsay2},  
the recent observations emphasize the importance of the 
wire surface or bulk structural imperfections on a length scale of the 
order of a micron or longer.  

Here, we present a detailed model for the current
irregularities formed in a current carrying wire with random
geometrical perturbations or bulk resistivity inhomogeneities. 
This
model enables not only the quantitative understanding of the
observed patterns and their origin, but also provides predictions
of electron transport properties in wires with various geometries
and crystalline structures. Together with further measurements
using the ultracold atomic probe, it is expected to shed new light
on electron transport, and to allow for less corrugated atomic
traps and guides to be developed for atom optics and quantum
technology\cite{folman}.

\begin{figure}
\includegraphics[width=\columnwidth]{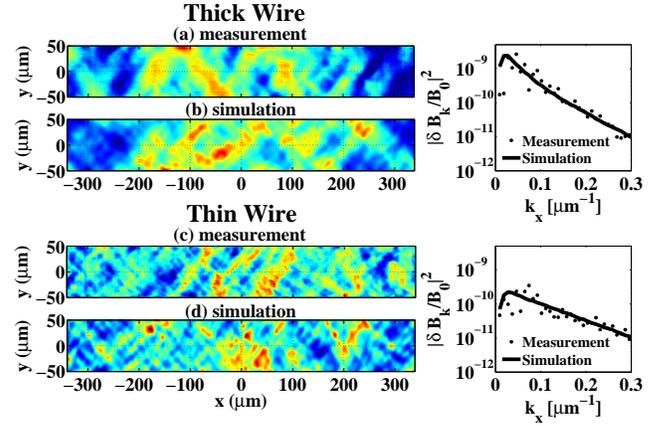}
\caption{Measured \cite{simon} and simulated patterns of atomic
density (normalized) $\sim$3.5$\mu$m above (a-b) a thick wire ($H=2\mu$m), and (c-d)
a thin wire ($H=280$nm), with grain size of $\sim 70$nm and $\sim
40$nm, respectively. 
Atomic density variations measure
corrugations in the magnetic field component $\delta B_x$ along
the wire axis $\hat{x}$. These corrugations are due to current
directional deviations from the main current along $\hat{x}$. The
trapping potential ensures a cloud width of $\sim 1\mu m$ along
$\hat{y}$ and $\hat{z}$, and hundreds of $\mu m$ in the $\hat{x}$
direction. To create the above 2d maps, the different 
transverse locations
are scanned across the wire at a constant height.  The simulation assumes
a combination of bulk resistivity inhomogeneity and geometrical 
perturbations of the wire with parameters chosen such that the 
power spectrum of the magnetic field along $\hat{x}$,
averaged over $\hat{y}$ and over many realizations of the simulation,
fits the measured power spectrum (right).    
 \label{fig1}}
\end{figure}

The results of our calculations are demonstrated in
Fig.~\ref{fig1}, comparing the measured atomic density patterns
with patterns calculated by assuming random imperfections
of the wire geometry or bulk resistivity. Each spectral (Fourier) component of these
imprefections is a plane wave with a
random phase and an amplitude taken from a non-white isotropic
power spectrum modeled in Ref.~\onlinecite{simon}. The angular
preference of the patterns emerges from a universal electron
scattering mechanism described below, where the apparent
difference in the spectral composition of the density fluctuations
above the thick wire of thickness 2$\mu$m (Fig.~\ref{fig1}a-b) and
the thin wire of thickness 280nm (Fig.~\ref{fig1}c-d) is
attributed to the different nature of the wire imperfections. 

\begin{figure}
\includegraphics[width=\columnwidth]{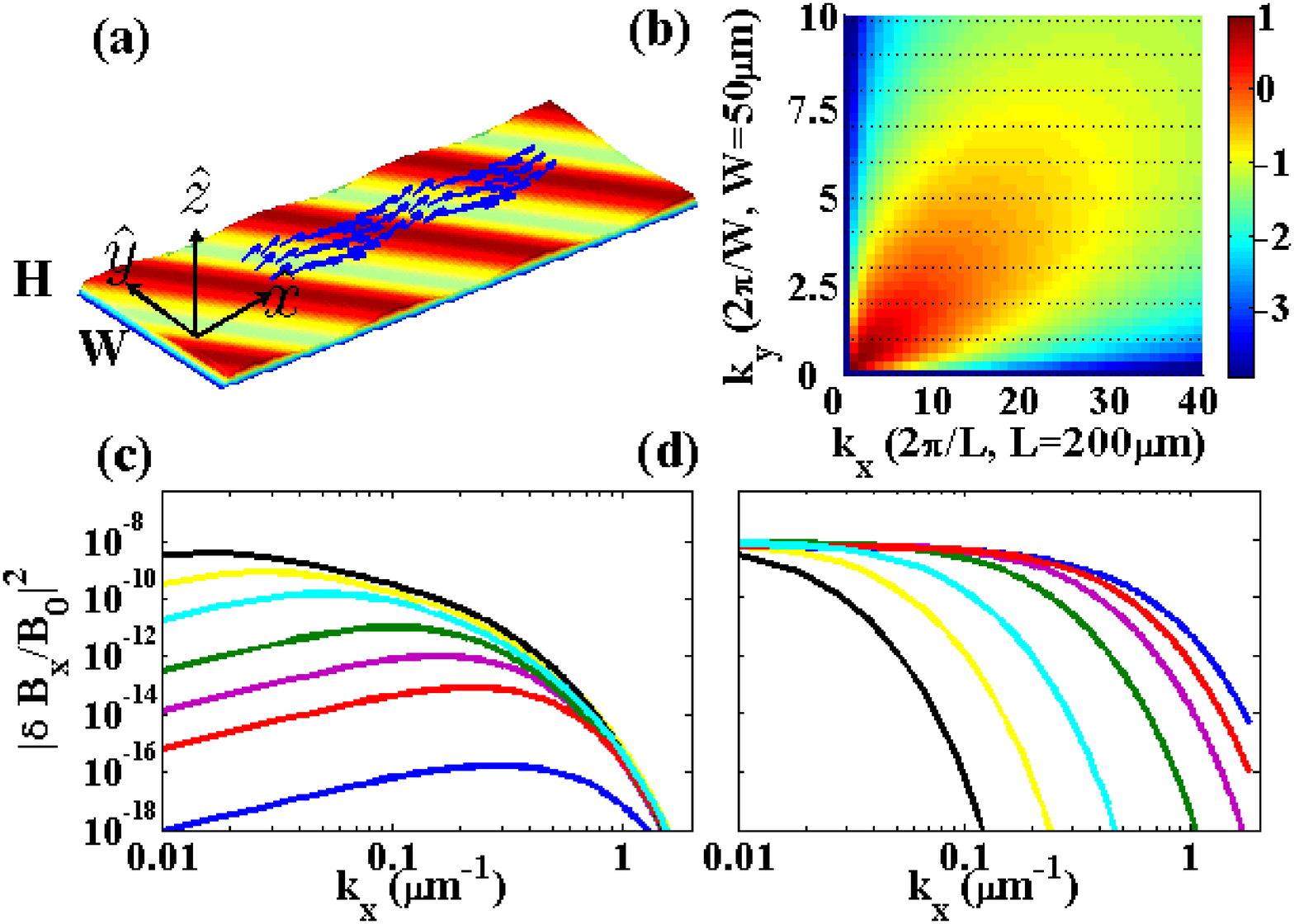}
\includegraphics[width=\columnwidth]{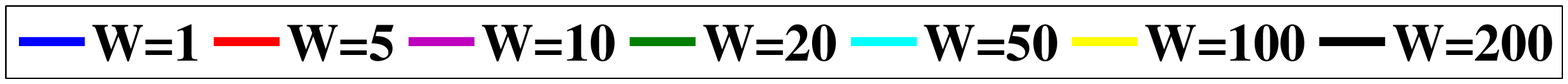}
\caption{(a) A single Fourier component (plane wave) of the planar
resistivity perturbations due to bulk inhomogeneity or wire
thickness variations induces current flow directional changes
(arrows). The current tilts along low resistivity wavefronts and
across high resistivity wavefronts. (b) The amplitudes of transverse
current component $\delta J_y(\kap)$ [$\kap\equiv (k_x,k_y)$],
generated by resistivity perturbations $\delta\rho(\kap)\propto
1/\kappa$ (see Ref.~ \onlinecite{simon}), are proportional to $\sin 2\theta_{\kap}$
(color scheme). For a wire with a finite length $L$ and width $W$,
these amplitudes are calculated at discrete values of $\kap$,
which are integer products of $2\pi/L$ and $2\pi/W$. (c) As a
consequence of this discreteness, the corrugations at long
wavelengths ($k_x W<1$) are suppressed when the wire becomes
narrower: the density of $k_y$ states becomes lower and
no corresponding values of $k_y$ exist along the maximum
scattering amplitude line $k_y \sim k_x$. Here, the power spectrum
of the magnetic field corrugations at 3.5$\mu$m above the center
of the wire is shown as a function of $k_x$ for resistivity
perturbations $\delta\rho_{\kap}/\rho_0=3.4\cdot 10^{-4} (\kappa_0/\kappa)$
with $\kappa_0=2\pi/680\mu$m$^{-1}$. (d) For comparison, the
magnetic field corrugations above the center of the wire are shown
for wires of different widths for a model assuming edge
fluctuations with $|\delta
y_{\pm}(k_x)|=10$nm$\times (\kappa_0/k_x)$]. Here, short wavelength
components are suppressed when the wire becomes wider. }
\label{fig2}
\end{figure}

Our calculations are performed for a metallic wire as in
Fig.~\ref{fig2}a, having a rectangular cross-section of width $W$
in the $\hat{\bf y}$ direction and thickness $H$ in the $\hat{\bf
z}$ direction, and carrying a current density ${\bf J}({\bf
r})=J_0\hat{\bf x}+\delta {\bf J}({\bf r})$. Here $J_0=I/WH$ is
the regular current density for a total current $I$. Utilizing
Ohm's law ${\bf J}={\bf E}/\rho$ where ${\bf E}$ is the electric
field, and Maxwell's equation $\nab\times{\bf E}=0$, we find
\begin{equation}
\nab\times {\bf J}=-\frac{{\nab}\rho}{\rho}\times {\bf J}\ ,
\label{eq:inhomog}
\end{equation}
with  $\rho=\rho_{0}+\delta \rho ({\bf r})$ being the isotropic
resistivity. Although significant resistivity perturbations may
exist in a polycrystalline metal near grain boundaries (length
scale of nanometers), one may safely assume that over most of the
length scales of interest (microns) $\delta\rho\ll\rho$, such that
in the Fourier expansion $\delta\rho({\bf r})= \sum_{\bf k}
\rho_{\bf k}e^{i{\bf k}\cdot{\bf r}}$ one has $|\rho_{\bf k}|\ll
\rho_0$ for any relevant wave number ${\bf k}\equiv (k_x,k_y,k_z)$.  
By keeping only terms up to
first order in the resistivity gradient ${\bf \nabla}\rho$ and
using the current continuity equation $\nab\cdot\delta{\bf J}=0$
we obtain the solution for the components of the current
irregularities as a function of the bulk inhomogeneity
\begin{equation}
\delta{\bf J}^{(\rm bulk)}({\bf k})=J_0\left(\frac{k_x}{|{\bf
k}|^2}{\bf k}-\hat{\bf x}\right) \frac{\delta\rho_{\bf
k}}{\rho_0}, \label{Jinh}
\end{equation}
where the transverse components $(k_y,k_z)$ of the wave
vector ${\bf k}$ take the discrete values $2\pi(m/W,n/H)$ with
integers $m$ and $n$, $-\infty<m,n<\infty$.

The horizontal transverse current irregularities $\delta
J_y^{({\rm bulk})}$ are proportional to $k_x k_y/k^2\propto  \sin
2\theta_{{\bf k}}$, where  $\theta_{{\bf k}}\equiv
\tan^{-1}(k_y/k_x)$ is the angle in the x-y plane.  This
immediately implies that transverse currents are predominantly
generated by Fourier components of the resistivity perturbations
with wavefronts oriented at $\pm 45^{\circ}$. Vertical current
irregularities $\delta J_z^{({\rm bulk})}$ are proportional to
$k_x k_z/k^2\propto \sin 2\phi_{\bf k}$, where $\phi_{\bf k}\equiv
\tan^{-1}(k_z/k_x)$. Vertical currents are therefore significant
only for Fourier components satisfying $k_x \sim k_z$ ($\phi_{\bf
k}\sim \pm 45^{\circ}$), namely, for longitudinal wavelengths
$2\pi/k_x$ of the order of the thickness $H$ or less,
corresponding to non-zero values of $k_z$. For thin wires, these
wavelengths are usually beyond the spatial measurement resolution
in the $x-y$ plane. At wavelengths of interest, much larger than $H$
($k_x \ll k_z$), vertical currents are suppressed as $\delta J_z^{({\rm
bulk})}\propto \frac{k_xH}{2\pi} \ll 1$.

In the following we refer to the spectral regime $(k_x,k_y)\ll
2\pi/H$ as the "thin film limit", where only contributions from
Fourier terms with $k_z=0$ are important.  We will then consider
the film as two-dimensional and characterize it by the real-space
vector $\xiv\equiv (x,y)$ and Fourier space vector $\kap\equiv
(k_x,k_y)$. Thickness variations of the wire $\delta H(x,y)$ may
then be regarded as irregularities of the thin film resistivity
$\delta \rho^{{\rm thickness}}=-\rho_0 \delta H/H$.
Figure~\ref{fig2}a demonstrates the generation of periodic
horizontal current directional deviations due to resistivity
perturbations originating from bulk or thickness variations.

A typical magnetic potential along an elongated trap, such as that used 
in Ref. \onlinecite{simon}, is determined mainly by the longitudinal 
component of the magnetic field fluctuations at the trapping position. 
Its Fourier spectrum at a height $z_0$ is related to the current 
irregularities in the wire by
\begin{eqnarray}
&&\delta B_x(k_x,k_y,z_0)=\frac{\mu_0}{2} \int_{-H}^{0} dz'\ e^{-\kappa |z_0-z'|}\times \nonumber \\
&&\times \left[\delta J_y(k_x,k_y,z') +i\sin\theta_{ \kap}\delta
J_z(k_x,k_y,z')\right]\ , \label{dBx_k}
\end{eqnarray}
where $\mu_0$ is the permeability of the vacuum. Here, the free
space (continuous) Fourier transformations $\delta {\bf
J}(k_x,k_y,z')$ may be approximated by their discrete form as in
Eq.~(\ref{Jinh}) if $\kappa W\gg 1$ and $z_0\ll W/2$. Substituting
$\delta{\bf J}$ of Eq.~(\ref{Jinh}) into this expression one finds
that $\delta B_x(\kap)\propto e^{-\kappa z_0}\sin 2\theta_{\kap}$
limiting the spatial resolution in the x-y plane by the
measurement distance $z_0$.

The $\sin 2\theta_{\kap}$ dependence together with a $\sim
1/\kappa$ dependence of the resistivity perturbations
(found in the spectral analysis of the data in Ref.~\onlinecite{simon}),
demonstrated by the color map in Fig.~\ref{fig2}b,
describe well the behavior in the continuum limit $W\rightarrow
\infty$. However, for finite widths (and a finite measurement
length $L$) $k_x$ and $k_y$ assume only discrete values which are
integer multiples of $2\pi/L$ and $2\pi/W$, respectively, as
demonstrated by the grid of dots superposed on the color map. It
follows that for small values of $k_x$, no counterparts $k_y$
exist on the grid which lie in the region where $|\sin 
2\theta_{\kap}|$ is large, or more specifically, around the line
$\theta_{\kappa}=45^{\circ}$. This implies that at wavelengths
larger than the wire width $W$ the current irregularities are
significantly suppressed beyond the suppression caused by the 
reduction of grid points. This prediction is demonstrated in the
power spectrum shown in Fig.~\ref{fig2}c. This result is very
different from the effect of current irregularities due to edge
roughness, which characterized measurements of atomic density
fluctuations in some previous works \cite{lukin,orsay1,orsay2}. In
that case, the short wavelengths are exponentially suppressed near
the center of the wire, while only wavelengths of the order of the
wire width or more are effective (Fig.~\ref{fig2}d).

\begin{figure}
\includegraphics[width=\columnwidth]{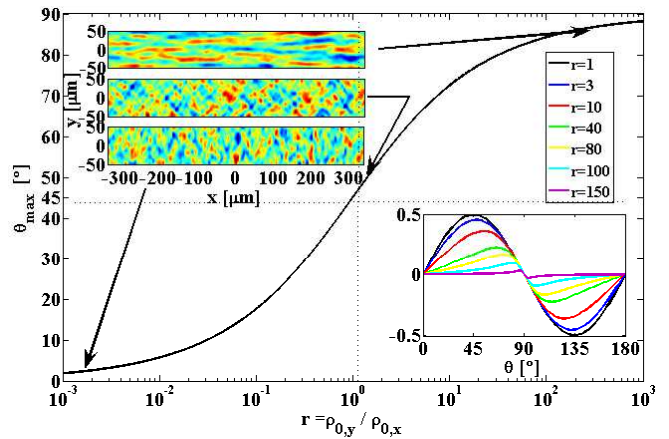}
\caption{For a current flowing through an electrically anisotropic wire, 
the perturbation wavefront angle $\theta_{\kap}^{\rm max}$,  giving rise to 
maximum transverse electron scattering, will depend on the ratio $r=\rho_y/\rho_x$ 
between the transverse
and longitudinal resistivities (main plot). The two-dimensional maps show the 
predicted atomic density above a wire similar to that presented in
Fig.~\ref{fig1} in the extreme cases $\theta_{\kap}=0^{\circ}$ ($r\ll 1$), 
$90^{\circ}$ ($r\gg 1$) and the isotropic case
$\theta_{\kap}=45^{\circ}$ ($r=1$). Bottom inset - 
the magnetic corrugation amplitude as a function of
$\theta_{\kap}$, which is suppressed when $r>1$. For details
see Ref.~\protect\onlinecite{tal}.} \label{fig3}
\end{figure}

Another prediction of our model is obtained when we generalize the
situation to the case where the conducting wire is electrically
anisotropic, such that the resistivity is a diagonal tensor and
Ohm's law generalizes to $E_j=\rho_j J_j$ for $j=x,y,z$. In this
case Eq.~(\ref{Jinh}) becomes \cite{tal}
\begin{equation}
\delta{\bf J}^{(\rm bulk)}({\bf k})=J_0\left(\frac{k_x}{{\bf
k}\cdot{\bf q}}{\bf q}-\hat{\bf x}\right) \frac{\delta\rho_{x,{\bf
k}}}{\rho_{x,0}}, \label{Jinh_aniso}
\end{equation}
where ${\bf q}=(k_x/\rho_x,k_y/\rho_y,k_z/\rho_z)$.  In the limit
of a thin film, where $k_z=0$, the horizontal transverse current
irregularities $\delta J_y$ are proportional to $\sin
2\theta_{\kap}/(1+(r-1)\cos^2 \theta_{\kap})$, where
$r=\rho_y/\rho_x$ is the resistivity ratio. As demonstrated in
Fig.~\ref{fig3}, the scattering at angles
$\theta_{\kap}<45^{\circ}$ is suppressed if $r>1$ and enhanced if
$r<1$, thus changing the preferred scattering wavefront angle in
the range $0^{\circ}<\theta_{\kap}<90^{\circ}$. The
overall magnetic corrugations are suppressed as $r^{-3/4}$ in the
limit of high anisotropy $r\gg 1$.

Now we turn to a more detailed theory of current irregularities 
due to geometrical imperfections of the wire \cite{orsay_comment}. We solve
Eq.~(\ref{eq:inhomog}) with $\delta\rho\rightarrow 0$ and with
boundary conditions ensuring that the current flows parallel to
the boundaries. Taking the upper and lower surfaces of the wire at
$z=\pm H/2+ \delta z_{\pm}$ and the right and left edges at $y=\pm
W/2+\delta y_{\pm}$, where $\delta z_{\pm}(x,y)$ and $\delta
y_{\pm}(x,z)$ are small fluctuations of the corresponding
surfaces, we obtain the following boundary conditions,
\begin{eqnarray}
\delta J^{}_{y}(x,\pm\frac{W}{2},z) &=& J^{}_{0}\frac{\partial
\delta y_{\pm}}{\partial x}\   , \nonumber \\
\delta J^{}_{z}(x,y,\pm\frac{H}{2})&=& J^{}_{0}\frac{\partial
\delta z_{\pm}}{\partial x}\ , \label{boundary}
\end{eqnarray}
where terms of second or higher orders in $\delta z_{\pm}$ and
$\delta y_{\pm}$ were omitted. The current irregularities are then
written as a sum of two terms,
\begin{eqnarray}
\delta {\bf J}^{(\rm surf)}({\bf r})&=& \sum_{k_x} e^{ik_x x}
\left[\delta{\bf J}^W_{k_x}(y,z)+\delta{\bf J}^H_{k_x}(y,z)\right]\ . \label{dJ}
\end{eqnarray}
Eq.~(\ref{eq:inhomog}) with ${\bf \nabla}\rho=0$ together with the
continuity equation ${\bf \nabla}\cdot{\bf J}=0$ imply that the
current can be written as the gradient of a potential function
$\delta {\bf J}^{(\rm surf)}={\bf \nabla}{\cal F}$, which
satisfies the Laplace equation $\nabla^2 {\cal F}=0$. It follows
that the terms in Eq.~(\ref{dJ}) have the form
\begin{eqnarray}
\delta{\bf J}^W_{k_x}(y,z) = ik_x J_0\sum_{n,\pm} {\bf
a}_{n,\pm}(k_x)e^{i2\pi nz/H}e^{-|y\mp W/2|/\lambda_n}
\label{JW} \\
\delta{\bf J}^H_{k_x}(y,z) = ik_x J_0\sum_{m,\pm} {\bf
b}_{m,\pm}(k_x)e^{i2\pi my/W}e^{-|z\mp H/2|/\lambda_m}, \label{JH}
\end{eqnarray}
where the exponential terms describe the attenuation of current
fluctuations induced by each boundary perturbation at a distance
$\lambda_n=[k_x^2+(2\pi n/H)^2]^{-1/2}$ from the left/right
boundaries and $\lambda_m=[k_x^2+(2\pi
m/W)^2]^{-1/2}$ from the top/bottom boundaries. Since $\delta {\bf
J}$ is derivable from a scalar function, it follows that each of
the vectorial coefficients ${\bf a}_{n,\pm}$ and ${\bf b}_{m,\pm}$
can be derived from the corresponding scalar coefficients. Linear
equations are obtained for these scalar coefficients when $\delta
{\bf J}$ of Eq.~(\ref{dJ}) is substituted in the boundary
conditions~(\ref{boundary}) \cite{unpublished}.

Next, we describe the solutions of these equations for a few
typical simple cases. The term $\delta {\bf J}^W_{k_x}$
(Eq.~(\ref{JW})) is significant when the edge perturbations
$\delta y_{\pm}$ are large and the measurement height is
comparable to the wire width $W$. This situation was discussed in
previous works \cite{lukin,orsay2}.
Here we concentrate on the other limit, where the field is
measured at a low height and a large distance from the edges
compared to $\lambda_n$ such that $\delta{\bf J}^W_{k_x}\sim 0$ for most
values of $k_x$. The surface height fluctuations $\delta
z^{\pm}_{\kap}$ then generate the following current irregularities
(Eq.~(\ref{JH}))
\begin{eqnarray}
\delta J_y^{\rm (surf)}({\bf r})&\approx &-J_0\sum_{\kap} e^{i\kap\cdot \xiv} k_x \frac{k_y}{\kappa}
\frac{\cosh(\kappa z)}{\sinh(\kappa H/2)} \frac{\delta
H_{\kap}}{2}\ ,  \label{JyA} \\
\delta J_z^{\rm (surf)}({\bf r})& \approx & J_0\sum_{\kap}e^{i{\kap}\cdot \xiv} ik_x
\frac{\cosh(\kappa z)}{\cosh(\kappa
H/2)}\delta z_{\kap}^{\rm mean}\ , \label{JzS}
\end{eqnarray}
where $\delta H_{\kap}=\delta z^+_{\kap}-\delta z^-_{\kap}$ are wire thickness
variations and $\delta z^{\rm mean}_{\kap}=(\delta z^+_{\kap} + \delta z^-_{\kap})/2$
are height fluctuations of the center of the wire. In the thin
film limit $\kappa H\ll 1$, we find $\kappa z\ll 1$ for $|z|\leq
H/2$ such that $\delta J_y$ assumes a form similar to
Eq.~(\ref{Jinh}) with $\delta \rho_{\kap}/\rho_0\rightarrow
-\delta H_{\kap}/H$
\begin{equation}
\delta J_y^{\rm (surf)}({\bf r})\approx
-J_0\sum_{\kap}e^{i\kap\cdot\xiv}\sin 2\theta_{\kap}
\frac{\delta H_{\kap}}{2H}\ . \label{JyA0}
\end{equation}
In the same limit, the magnitude of the vertical current component
$\delta J_z$ becomes negligible, since $\delta J_z^{({\rm
surf})}\propto k_x \delta z^{\rm mean}_{\kap}= k_x H(\delta
z_{\kap}^{\rm mean}/H)\ll \delta z_{\kap}^{\rm mean}/H$.

Equation~(\ref{JyA0}) shows that the surface roughness $\delta z_{\pm}$ causes current 
irregularities mainly through the thickness variations $\delta H=\delta z_+
-\delta z_-$. In thin films, long scale surface height variations $\delta z_+$, 
which follow bottom (wafer) surface variations $\delta z_-$ ,  are not expected to 
cause significant current variations.  On the other hand, the effect of thickness 
variations is expected to be more pronounced
in thin films, where $\delta H/H$ is larger than in thicker wires. The power spectrum 
of the measured magnetic field pattern above the thin wire  (Fig.~\ref{fig1}c), which 
was analyzed in Ref.~\onlinecite{simon}, could be explained by
a model assuming that thickness variations exist mainly at short 
length scales (below $\sim$20$\mu$m), while $\delta z_+\approx \delta z_-$ at long length
scales. This may explain the shorter characteristic length scale of the features
in the thin wire (Fig.~\ref{fig1}c) relative to the thick wire (Fig.~\ref{fig1}a), where  the 
measured surface roughness was not sufficiently large to account for the 
magnetic field fluctuations even if the bottom surface variations were assumed 
to be uncorrelated with the top surface. This analysis implies that
bulk resistivity perturbations with $\propto 1/k$ spectrum could play an 
important role in the thick wire, but they are much smaller in the thin wire. 

To conclude, we have presented a detailed model for current directional deviations
in thin wires with random imperfections on a length scale of the order of 
a micron or longer.  
These deviations may arise either from bulk
resistivity inhomogeneities or from geometrical perturbations of the wire, 
where the significance of each factor depends on the wire thickness and 
fabrication process. In both cases, electron scattering
is dominant at wavefronts oriented at $\pm 45^{\circ}$ relative to
the main current axis. The model predicts a strong suppression
of long wavelength current deviations originating from bulk or
surface corrugations in narrower wires. Electrically anisotropic
materials are also capable of significantly suppressing these
deviations. Such analysis opens the road for material
engineering to considerably improve atom optics on atomchips where
currents are used for creating magnetic potentials for atom trapping and guiding. 
Comparison of this theory with further cold
atom magnetometry or other measurements providing high field sensitivities
and spatial resolution, will enable deeper understanding of
electron transport in thin films.


We thank the team of the Ben-Gurion University Weiss Family
Laboratory for Nanoscale Systems (www.bgu.ac.il/nanofabrication)
for the fabrication of the chip and J\"urgen Jopp of the
Ben-Gurion University Ilse Katz Center for Nanoscale Science for
assisting with surface measurements. R.F. thanks Yoseph (Joe)
Imry. We acknowledge support by the FWF, the DFG, the German
Ministry of Education and Research (DIP), the EC 'atomchip' (RTN)
consortium, the American-Israeli Foundation (BSF) and the Israeli
Science Foundation.


\begin{thebibliography}{999}
\bibitem{orsay_comment} Our treatment gives a full solution for small arbitrary 
fluctuations in a wire with rectangular cross section. 
The case of top surface
fluctuations was given in Ref.~\protect\onlinecite{orsay2} only for symmetric
fluctuations that contribute to the magnetic field above the
middle of the wire.
\end{thebibliography}
\end{document}